\documentclass[12pt]{article}
\usepackage{graphicx}
\usepackage{epsfig}
\begin{document}
\title{Isgur-Wise function in a QCD potential model with coulombic potential as perturbation}
\author{$^{1}$Bhaskar Jyoti Hazarika, $^{2}$ Krishna Kingkar Pathak and $^{3}$ D K Choudhury \\
$^{1}$Dept of Physics,Pandu College,Guwahati-781012,India\\
$^{2}$ Dept of Physics,Arya Vidyapeeth College,Guwahati-781016,India\\
$^{3}$Dept. of Physics, Gauhati University, Guwahati-781014,India}
\date{}
\maketitle
\begin{abstract}
We study heavy light mesons in a QCD inspired quark model with  the Cornell potential$-\frac{4\alpha _{S}}{3r}+br+c$. Here we consider the linear term $br$ as the parent and  $-\frac{4\alpha _{S}}{3r}+c$ i.e.the Coloumbic part as the perturbation.The linear parent leads to Airy function as the unperturbed wavefunction.We then use the Dalgarno method of perturbation theory to obtain the total wavefunction corrected upto first order with Coulombic piece as the perturbation.With these wavefunctions , we study the Isgur-Wise function and calculate its slope and curvature. 
\end{abstract}
\section{Introduction}

Considerable efforts have been made in understanding the physics of hadrons containing at least one heavy quark since long [1-9].It is well known that the heavy quark symmetry in the heavy quark limit leads to a single form factor called as Isgur-Wise(I-W) function which can describe the heavy quark bilenear current matrix elements of weak decay.The basic ingradient of the I-W function is the hadronic wave function,the determination of which becomes such a crucial factor.The potential models for this purpose is quite helpful as they contain more input parameters and hence has  its firm basis.\\

Under such circumstances the I-W function has been investigated [3-9] with considerable success of valid degrees in different models. In the potential models `Cornell potential' is found to be more useful than the others. It leaves two options of choosing the parent (1) the Coulombic part  $-\frac{4\alpha _{S}}{3r}$ and (2) the linear potential br.\\

The slope and curvature of I-W function with the coulombic potential as the parent has already been reported for different heavy-light flavour mesons [10-15], which however had certain limitations.In coulombic potential as parent and linear as perturbation, the value of slope ( $\rho^{2}$) and curvature ($C$) was found to be too large in $\overline{MS}$ -scheme. Imposing V-scheme [14-18], with larger $\alpha_{s}$ the values were found to be improved [13] but still larger than the expectations. As an alternate approach,in the present work we choose linear term $`br'$ as the parent and Coulombic piece as the perturbation in finding the wave function.\\

As usual,two body  Schr\"{o}dinger equation is used with the ground state Fermi-Breit Hamiltonian in the absence of contact term and with Coulombic perturbation ,the wavefunction corrected upto first order is obtained by using the Dalgarno method \cite{10,19}. The relavistic effect is incorporated by using standard Dirac modification\cite{20,21} in a parameter free way.These wavefunctions are used  in the calculation of slope and curvature of Isgur -Wise function.\\

The rest of the paper is organised as follows : section 2 contains the formalism , section 3 the result and section 4 the conclusion and discussion.\\

\section{Formalism}
\subsection{The wavefunction}
We start with the ground state ($l=0$) spin independent Fermi-Breit Hamiltonian without the contact term given by \cite{10,11}:\\

 \begin{equation}                                                              
  H=-\frac{\nabla^{2}}{2\mu}-\frac{4\alpha_s}{3r}+br+c
\end{equation}
so that
\begin{equation}
H^{\prime}=-\frac{4\alpha_s}{3r}+c
\end{equation}
can be treated as perturbation to the unperturbed Hamiltonian :\\

\begin{equation}
H_{o}=-\frac{\nabla^{2}}{2\mu}+br
\end{equation}
In eq.(1), the strong coupling constant connected to the potential is a function of the momentum as
\begin{equation}
\alpha_{s}\left(\mu^{2}\right)=\frac{4\pi}{\left(11-\frac{2n_{f}}{3}\right)ln\left(\frac{\mu^{2}}{\Lambda^{2}}\right)}
\end{equation}
where $n_{f}$ is the number of flavour.
The constant $`c'$ at its natural scale is taken to be 1 GeV \cite{13}
The two body non relativistic Shr\"oedinger wave equation can be recasted as
\begin{equation}
H|\psi>=\left(H_{o}+H^{\prime}\right)|\psi>=E|\psi>
\end{equation}
The unperturbed wave function coresponding to $H_{0}$ are the Airy functions which after normalization can be written as :
\begin{equation}
\psi_{n}^{(0)}\left(r\right)=\frac{N}{2\sqrt{\pi}r}Ai(\left(2\mu b\right)^{\frac{1}{3}}+\rho_{0n})
\end{equation}

where $\rho_{0n}$ are the zeros of the Airy function $Ai\left(\rho_{0n}\right)=0$, $n=1,2,3..$ represent the principal quantum no.(of course for the ground state n=1) and N is the normalisation constant.\\

The $\rho_{0n}$ are given as \cite{22,23}:\\

\begin{equation}
\rho_{0n}=-[\frac{3\pi\left(4n-1\right)}{8}]^{\frac{2}{3}}
\end{equation}

The first order corection to wave function $\psi_{n}^{(1)}$ and energy $W_{n}^{(1)}$ are respectively given by 
\begin{equation}
H_{0}\psi_{n}^{(1)}+H^{\prime}\psi_{n}^{0}=W_{n}^{0}\psi^{1}+W_{n}^{1}\psi^{0}
\end{equation}
where $W_{n}^{0}$ is the unperturbed energy given as \cite{22} 
\begin{equation}
W_{n}^{0}=E_{n}= -\left(\frac{b^{2}}{2\mu}\right)^{\frac{1}{3}} \rho_{0n}
\end{equation}
and
\begin{equation}
W_{n}^{(1)}=\int_{0}^{+\infty}r^{2} H^{\prime}\left|\psi^{(0)}\left(r\right)\right|^{2} dr
\end{equation}
Since we consider the ground state ($n=1$ ) , so we drop the '$n$' from $ W_{n}^{0}$,$W_{n}^{(1)}$, $\psi_{n}^{(0)}$ and $\psi_{n}^{1}$.
The first order correction is :
\begin{equation}
 \psi^{1}\left(r\right)=-\frac{4\alpha_{s}}{3}\left(\frac{a_{0}}{r}+a_{1}+a_{2} r\right)
\end{equation}
As Airy function Ai(r) involve infinite series in $r$, so in calculating the coefficients $a_{0}$,$a_{1}$ and $a_{2}$ we have considered upto order $r^{3}$ and  are given by :\\

\begin{equation}
a_{0}=\frac{0.8808 \left(b\mu\right)^{\frac{1}{3}}}{\left(E-c\right)}-\frac{a_{2}}{\mu\left(E-c\right)}+\frac{4 W^{1}\times 0.21005}{3\alpha_{s}\left(E-c\right)}
\end{equation}
\begin{equation}
 a_{1}=\frac{ba_{0}}{\left(E-c\right)}+\frac{4\times W^{1}\times0.8808\times \left(b\mu\right)^{\frac{1}{3}}}{3\alpha_{s}\left(E-c\right)}-\frac{0.6535\times \left(b\mu\right)^{\frac{2}{3}}}{\left(E-c\right)}
\end{equation}
\begin{equation}
a_{2}=\frac{4\mu W^{1}\times0.1183}{3\alpha_{s}}
\end{equation}
The total wave function corrected upto first order with normalisation is
\begin{eqnarray}
\psi_{coul}\left(r\right)&=&\psi^{\left(0\right)}\left(r\right)+\psi^{\left(1\right)}\left(r\right)\\&=&\frac{N_{1}}{2\sqrt \pi}\left[\frac{Ai(\left(2\mu b\right)^{\frac{1}{3}}+\rho_{01})}{r}-\frac{4\alpha_{s}}{3}\left(\frac{a_{0}}{r}+a_{1}+a_{2}r\right)\right]
\end{eqnarray}

where $ N_{1}$ is the normalisation constant for the total wave function $\psi_{coul}\left(r\right)$ with subscript 'coul' means coulombic potential as perturbation.\\

The relativistic version of eq (17)is obtained by multiplying it with $\left(\frac{r}{a_{Bohr}}\right)^{-\epsilon}$.\\

$a_{Bohr}$  depends on $\alpha_{s}$ as :\\

\begin{equation}
a_{Bohr}=\frac{3}{4\mu \alpha_{s}}
\end{equation} 
and
\begin{equation}
\epsilon =1-\sqrt{1-\left(\frac{4\alpha_{s}}{3}\right)^{2}}
\end{equation}
Thus,relativistic wavefunction is :
\begin{equation}
\psi_{rel}\left(r\right)=\psi_{coul}\left(r\right)\left(\frac{r}{a_{Bohr}}\right)^{-\epsilon}
\end{equation}
\subsection{Isgur-Wise function}
The Isgur-Wise function is written as \cite{1,2} :
\begin{eqnarray}
\xi\left(v_{\mu}.v^{\prime}_{\mu}\right)\nonumber&=&\xi\left(y\right)\\&=&1-\rho^{2}\left(y-1\right)+ C\left(y-1\right)^{2}+...
\end{eqnarray}
where 
\begin{equation}
y= v_{\mu}.v^{\prime}_{\mu}
\end{equation}
and $v_{\mu}$ and $v^{\prime}_{\mu}$  being the four velocity of the heavy meson before and after the decay.The quantity $\rho^{2}$  is the slope of I-W function at $y=1$ and known as charge radius :\\

\begin{equation}
\rho^{2}= \left. \frac{\partial \xi}{\partial y}\right.|_{y=1}
\end{equation}
The second order derivative is the curvature of the I-W function known as convexity parameter :\\

\begin{equation}
C=\left .\frac{1}{2}\right. \left(\frac{\partial^2 \xi}{\partial y^{2}}\right)|_{y=1}
\end{equation}
For the heavy-light flavor mesons the I-W function can also be written as \cite{6,11} :\\

\begin{equation}
\xi\left(y\right)=\int_{0}^{+\infty} 4\pi r^{2}\left|\psi\left(r\right)\right|^{2}\cos pr dr
\end{equation}
where\\

\begin{equation}
p^{2}=2\mu^{2}\left(y-1\right)
\end{equation}

the wavefunction  eq.(19) with relativistic effect is used in the calculation of $\xi\left(y\right)$ given by eq.(24).\\

\section{Calculation and Results}

We have calculated   the  values of charge radius and convexity parameter of the I-W function  given by  eq.(20) for two set of coupling constants both in $\overline{MS}$ and $V$ -scheme [14-17]. \\

Regarding the use of the above mentioned schemes \cite{11,12} we note that with  $n_{f}=4$ and $n_{f}=5$ and fixing $\Lambda_{QCD}=0.216 GeV$ [24]the coresponding value of $\alpha_{\overline{MS}}$ at the scale of 1.5 GeV  and 8 GeV are respectively $0.39$ and  $0.22$ \cite{24} . The respective change of $\alpha_{\overline{MS}}$ to  $\alpha_{v}\left(\frac{1}{r^{2}}\right)$ in the V  scheme\cite{14,15,16,17} for three different choices of scale $\overline{\mu}$ are  calculated \cite{12} and shown in the table 1 here.Although there is no fundamental reason for the choice ,we have chosen the two renormalization schemes ($\overline{MS}$  and $V$-schemes) to fecilitate the comparision of our result with the previous work[12,13]. Also we use the same model parameter $b=0.183GeV^{2}$ from charmonium spectroscopy[25,26] \\

 For these calculations,we have used the expressions for $E,W^{1}$ ,$a_{0}$,$a_{1}$ given by equations (10),(12),(13),(14) respectively.These are shown in the table 2 and table 3.The result of $\rho^{2}$ and $c$ in the present work is shown in table 4.  We also compare the present result with that of previous work with linear as the perturbation \cite{13} in $V$-scheme which was an improvement over $\overline{MS}$ -scheme and is shown in table 4. \\

In table 4, we give a list of predictions of $\rho^{2}$ and $C$ in different theoretical models.\\

In evaluating the various integrations, we use numerical method of integration in mathmatica software. \\

\begin{table}
\begin{center}
\caption{ The value of $\alpha_{v}\left(\frac{1}{r^{2}}\right)$ for different choices of $\overline{\mu}$ .}
\begin{tabular}{|c|c|c|c|}\hline
choices&$\overline{\mu}=\frac{1}{r}$ & $\overline{\mu}=\frac{e^{-\gamma_{E}}}{r}$ & $\overline{\mu}=\frac{e^{-\gamma_{E}-\frac{a_{1}}{2\beta_{0}}}}{r}$ \\\hline
$\alpha_{\overline{MS}}\left(m_{b}\right)=0.22,n_{f}=5$&0.259 &0.261&0.258\\\hline
$\alpha_{\overline{MS}}\left(m_{c}\right)=0.39,n_{f}=4$&0.693 &0.651&0.604\\\hline

\end{tabular}
\end{center}
\end{table}

\begin{table}
\begin{center}

\caption{The  values of $W^{1}$ and  E in GeV}

\begin{tabular}{|c|c|c|c|}\hline
Mesons&$E$&\multicolumn{2}{c|}{$W^{1}$}\\\cline{3-4}
 & &\multicolumn{1}{l|}{$\overline{MS}$ scheme}&V scheme\\\hline
$D$&0.3898&0.0467 &0.08314\\\hline
$D_{s}$&0.4291&0.5137 &0.0915\\\hline
$B$&0.4072& 0.02742&0.0327\\\hline
$B_{s}$&0.4553& 0.0308&0.0366\\\hline

\end{tabular}
\end{center}
\end{table}

\begin{table}
\begin{center}

\caption{List of $a_{0}$,$a_{1}$ and $a_{2}$  }

\begin{tabular}{|c| c| c| c| c| c| c|}\hline
Mesons&\multicolumn{2}{c|}{$a_{0}$}&\multicolumn{2}{c|}{$a_{1}(GeV)$}&\multicolumn{2}{c|}{$a_{2}(GeV^{2})$}\\\cline{2-7}
 &\multicolumn{1}{l|}{V scheme}&$\overline{MS}$ scheme&\multicolumn{1}{l|}{V scheme}&$\overline{MS}$ scheme &\multicolumn{1}{l|}{V  scheme}&$\overline{MS}$ scheme\\\hline
$D$&0.2143&0.1943 &-0.006138&-0.007877 &0.00293&0.002933\\\hline
$D_{s}$&0.238&0.21387 &-0.00916&-0.01257 &0.0043&0.0043036\\\hline
$B$&0.2245&0.2029 &-0.00749&-0.0099 &0.00349&0.00348\\\hline
$B_{s}$&0.254&0.2269 &-0.0114&-0.01604 &0.005446&0.00547 \\\hline

\end{tabular}
\end{center}
\end{table}

\begin{table}
\begin{center}

\caption{Values of $\rho^{2}$ and $C$  in our work and its comparision to other work }

\begin{tabular}{c c c c}\hline
 &Our work &  & \\\hline
Scheme&Mesons&$\rho^{2}$&$C$\\\hline
 $\overline{MS}$-scheme&$D$&0.7936 &0.0008\\
 &$D_{s}$&1.186 &0.002\\
 &$B$&0.89 &0.0004\\
 &$B_{s}$&1.41 &0.0012\\\hline
$V$- scheme&$D$&0.896 &0.00306\\
 &$D_{s}$&1.352 & 0.0077\\
 &$B$&0.912 &0.0007\\
 &$B_{s}$&1.421 &0.00155\\\hline
 &Other work & & \\\hline
Previous work\cite{12,13}&$D$&1.136&5.377\\
 &$D_{s}$&1.083&3.583\\
 &$B$&128.28&5212\\
 &$B_{s}$&112.759&4841\\\hline
Le Youanc et al \cite{27}& &$\ge 0.75$&..\\
Le Youanc et al \cite{28}& &$\ge 0.75$&$\ge 0.47$\\ 
Rosner \cite{29}& &1.66&2.76\\
Mannel \cite{30,31}& &0.98&0.98\\
Pole Ansatz \cite{32}& &1.42&2.71\\
Ebert et al \cite{36}& &1.04&1.36\\
Simple Quark Model \cite{3}& &1&1.11\\
Skryme Model \cite{35}& &1.3&0.85\\
QCD Sum Rule \cite{34}& &0.65&0.47\\
Relativistic Three Quark Model \cite{4}& &1.35&1.75\\
Neubert \cite{33}& &0.82$\pm$0.09&..\\\hline
\end{tabular}
\end{center}
\end{table}

\section{Discussion and Conclusion}
Our calculated values of slope of I-W function in this work are found to be in good agreement with the other theoretical results (table 4). The lattice QCD evaluation of $\rho^{2}=0.83_{-11-22}^{+15+24}$ for B meson[37] and the experimental values of D meson $\rho_{D}^{2}=0.76\pm0.16\pm0.08$ [38] and $\rho_{D}^{2}=0.69\pm0.14$ [39] are also in good agreement with our calculated results.However, the values of C for each meson are found to be smaller in comparison to other theoritical values. The reason may be presumably due to  the cut off of the infinite series of Ai(z) upto $O(r^{3})$ as noted earlier and still such  small values can be considered as a success particularly for the B sector mesons as these values were very large in case of coulombic potential as parent[11-14].\\

This study of the Isgur-Wise function with  Coulombic part as perturbation shows a different picture as compared to the earlier work \cite{11,12,13}. With linear part as perturbation, the slope and curvature decrease with the  increase of  $\alpha_{s}$; while in this work , we have observed a reverse effect. Further,this analysis shows a great reduction in the values of $\rho^{2}$ and $C$ for all the mesons as compared to the previous work with linear part as perturbation. \\
Let us conclude the section with a few comments.\\
The strong coupling constant entering the coulombic potential is a function of the momentum in full QCD.But in potential model,it is nothing but a mere  parameter. Here we have used the strong coupling constant in the $\overline{MS}$ and $V$-scheme to fecilitate a proper comparision with the previous work with linear part as perturbation[12,13].\\
However,instead of using a particular renormalization scheme we could as well have considered the strong coupling constant merely as a free parameter in  the potential model to be fitted from data. Such a possibility is currently under study.\\

\end{document}